\documentclass[%
 aip,
 amsmath,amssymb,
reprint,%
]{revtex4-1}

\usepackage{graphicx}
\usepackage{dcolumn}
\usepackage{bm}
\usepackage[mathlines]{lineno}

\usepackage[utf8]{inputenc}
\usepackage[T1]{fontenc}
\usepackage{mathptmx}
\usepackage{etoolbox}

\makeatletter
\def\@email#1#2{%
 \endgroup
 \patchcmd{\titleblock@produce}
  {\frontmatter@RRAPformat}
  {\frontmatter@RRAPformat{\produce@RRAP{*#1\href{mailto:#2}{#2}}}\frontmatter@RRAPformat}
  {}{}
}%
\makeatother
\begin{document}

\title{Self-similar solutions for resistive diffusion, Ohmic heating and Ettingshausen effects in plasmas of arbitrary $\beta$}
\author{G. Farrow}
\affiliation{The Centre for Inertial Fusion Studies, The Blackett Laboratory, Imperial College, London SW7 2AZ, United Kingdom}
\affiliation{The Blackett Laboratory, Imperial College, London SW7 2AZ, United Kingdom}
\email{gf715@ic.ac.uk}
\author{G. Kagan}%
\affiliation{The Blackett Laboratory, Imperial College, London SW7 2AZ, United Kingdom}
\author{J. P. Chittenden}
\affiliation{The Centre for Inertial Fusion Studies, The Blackett Laboratory, Imperial College, London SW7 2AZ, United Kingdom}
\affiliation{The Blackett Laboratory, Imperial College, London SW7 2AZ, United Kingdom}

\date{\today}

\begin{abstract}
Magneto-inertial fusion (MIF) approaches, such as the MagLIF experiment, use magnetic fields in dense plasma to suppress cross-field thermal conduction, attempting to reduce heat losses and trap alpha particles to achieve ignition. However, the magnetic field can introduce other transport effects, some of which are deleterious. An understanding of these processes is thus crucial for accurate modelling of MIF. We generalise past work exploiting self-similar solutions to describe transport processes in planar geometry and compare the model to the radiation-magnetohydrodynamics code Chimera. We solve the 1D extended magnetohydrodynamic (MHD) equations under pressure balance, making no assumptions about the ratio of magnetic and thermal pressures in the plasma. The resulting ordinary differential equation (ODE) boundary value problem is solved using a shooting method, combining an implicit ODE solver and a Newton-Raphson root finder. We show that the Nernst effect dominates over resistive diffusion in high $\beta$ plasma, but its significance is reduced as the $\beta$ decreases. On the other hand, we find that Ettingshausen and Ohmic heating effects are dominant in low $\beta$ plasma, and can be observable in even order unity $\beta$ plasma, though in the presence of a strong temperature gradient heat conduction remains dominant. We then present a test problem for the Ohmic heating and Ettingshausen effects which will be useful to validate codes modelling these effects. We also observe that the Ettingshausen effect plays a role in preventing temperature separation when Ohmic heating is strong. Neglecting this term may lead to overestimates for the electron temperature at a vacuum-plasma interface, such as at the edge of a z-pinch. The model developed can be used to provide test problems with arbitrary boundary conditions for magnetohydrodynamics codes, with the ability to freely switch on terms to compare their individual implementations.
\end{abstract}

\maketitle

\section{Introduction}
\label{sec:intro}
Magneto-inertial fusion is an approach to controlled thermonuclear fusion that attempts to reduce the ignition criteria of conventional inertial confinement fusion experiments through the use of magnetic fields \cite{Lindemuth_1981, Lindemuth_1983, Thio_2008}. Even a small magnetic field can be amplified by a compressing plasma by a factor of $10^3$ or more \cite{Gotchev_2009}. These large magnetic fields suppress cross-field thermal conduction to reduce heat losses. They also act to trap alpha particles, improving energy deposition. However, they can also give rise to other transport effects beyond thermal conduction, some of which are deleterious, such as the Nernst effect which reduces ideal MHD flux compression \cite{braginskii}. In addition, the magnetic field can significantly complicate the fusion alpha particle transport \cite{appelbe_2019,appelbe_2021}, affect hydrodynamic and laser-plasma instabilities \cite{bissell_2010, strozzi_2015, garcia_rubio_2021, walsh_2019, Walsh_2020, watkins_2018} and complicate the interpretation of diagnostics \cite{walsh_prl_2017, schmit_2014, sio_2021, hansen_2020}. Thus, accurate modelling of the magnetic field and magnetised plasma dynamics is crucial to both predictions and understanding of experiments. A detailed understanding of plasma transport processes in a wide range of parameter regimes is a key component of this. 

There are many MIF schemes currently operating, such as the MagLIF experiment at Sandia National Laboratories \cite{Slutz_2010}, where pulsed power is used to implode a beryllium liner onto a laser pre-heated and axially pre-magnetised column of deuterium. MagLIF has observed neutron yields of more than $10^{13}$, but flux compression has been hampered by the Nernst effect \cite{slutz_2018, gomez_2020}. Recent ``mini-MagLIF'' experiments on OMEGA have attempted to use laser-driven cylindrical implosions to assist with this effort \cite{Hansen_2018}. Currently, research has begun on whether magnetic fields can lead to ignition on indirect-drive ICF implosions on the National Ignition Facility \cite{perkins_2017}.

An important parameter for understanding magnetised transport is the plasma $\beta$, the ratio of the thermal to magnetic pressures. This is because it has been shown that the ratio of the Nernst to resistive diffusion terms in the 1D induction equation is proportional to $\beta L_B/L_T$ in the magnetised limit, where $L_B$ and $L_T$ are the magnetic field and temperature scale lengths respectively \cite{Velikovich_2014}. The ratio of the heat conduction to the Ettingshausen effect in the energy balance equation has the same scaling. Thus the plasma $\beta$ can be used to estimate which transport terms are negligible, allowing the MHD equations to be simplified.  Past work has also studied how these effects depend on the magnetic Lewis number, the ratio of thermal to magnetic diffusivities \cite{Velikovich_2014}. However, it should be noted that the transport is gradient-driven, and thus depends on the imposed boundary conditions. We explore this further in this work.

Past theoretical work has focused on the high $\beta$ limit, as this is of interest for the large temperatures and densities in ignition-scale MIF experiments. This reduces the pressure balance to purely thermal pressure and allows the magnetic diffusion, Ettingshausen and Ohmic heating effects to be dropped, simplifying the problem. Velikovich \textit{et al.} used self-similar solutions to estimate flux and energy losses from hot D$_2$ plasma to a fixed, cold liner \cite{Velikovich_2014} and found that the profiles are significantly affected by the inclusion of the Nernst effect. A similar study by Garcia-Rubio \textit{et al.}, also in the high $\beta$ limit, included mass-ablation effects to estimate concentration gradient losses \cite{Garcia-Rubio_20181}. Both approaches were later generalised to the low $\beta$ regime \cite{Giuliani_2018} and \cite{Garcia-Rubio_20182}. However, the maintenance of a rigid wall constraint in these works prevents the interpenetration of plasma and magnetic field and significantly restricts the boundary conditions. For example, the wall must be treated as a cold temperature sink, meaning the effect of a finite difference in temperature at the two boundaries cannot be assessed. A recent generalisation by Velikovich \textit{et al.} removed this rigid wall constraint in the high $\beta$ limit, which allows for a finite density and temperature difference between the liner and the plasma \cite{Velikovich_2019}. The lower $\beta$ regime is however accessed in pulsed power experiments, which are of importance to laboratory astrophysics as well as magnetised target fusion schemes \cite{Lebedev_2019, sinars_2020, Haines_2011}. Furthermore, all magnetic confinement devices, including z-pinches, are in a low $\beta$ regime at the edge. Similarly, the extreme magnetic fields generated in magnetic flux compression schemes \cite{flux_compression_talk} mean that the magnetic pressure is quite substantial. In the earliest work on this, Garanin explored the zero $\beta$ regime by considering the diffusion of magnetic field into a dense plasma \cite{garanin}, finding that the temperature diverged on the vacuum-plasma interface. This work differs from others by using a Lagrangian form of the governing equations and thus a different self-similar variable, as we highlight in section \ref{sec:theory} and in the appendix \ref{sec:appendix_lagrangian}. We have found that the solutions produced using the two different self-similar variables are identical. Garanin's work is difficult to generalise due to the use of asymptotic expansions on the vacuum-plasma interface however, so we mostly proceed with the Eulerian form of the equations. We have also found that this yields improved numerical stability.

We further generalise past work to arbitrary $\beta$ and consider the transport of heat, density and magnetic field across the full domain. This means that our problem is more general in boundary conditions and parameter regimes. We solve for the subsonic evolution of a plasma with a discontinuity in the initial conditions. The purpose of this work is to understand how the dominance of the Nernst and thermal conduction effects in high $\beta$ plasmas changes as we move to a low $\beta$ parameter regime. We particularly focus on the Ettingshausen effect, providing a demonstration of its behaviour in electron-ion temperature separation. The solutions produced in this work can also be used to provide validation for MHD codes attempting to model magnetised transport effects. We believe that our solutions will be particularly useful in studies of low density plasma, as well as regimes where resistive MHD is a commonly used modelling tool. Examples of validation against the radiation-MHD code Chimera \cite{Chittenden_2004, ciardi_2007} are shown later.

This paper is structured as follows. In section \ref{sec:theory}, we discuss the governing equations of the problem to be solved and the self-similar technique used in this paper. Then in section \ref{sec:method}, we discuss the shooting method used to solve these equations and the MHD code used for comparison. Finally, in section \ref{sec:results}, we present a scan over plasma $\beta$ to elucidate the role of these additional transport effects. We present a robust test problem for the Ohmic heating and Ettingshausen effects, which may be of use to the pulsed-power community in particular. In section \ref{sec:temp_sep}, we demonstrate how the Ettingshausen can play a vital role in controlling electron-ion temperature separation in the low $\beta$ regime. 

\section{Theory}
\label{sec:theory}
\subsection{Problem geometry and setup}

\begin{figure}
    \centering
    \includegraphics[width = 0.45\textwidth]{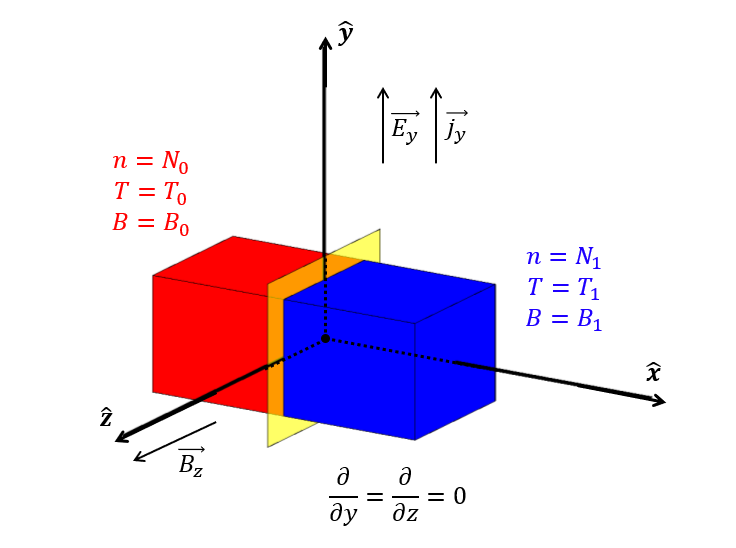}
    \caption{A schematic showing the geometry of our problem. The red spot represents the position of the original interface (at $x = 0$).}
    \label{fig:geometry_diagram}
\end{figure}

We model the problem in 1D planar geometry, as in past work \cite{Velikovich_2014, Velikovich_2019, Giuliani_2018}. This is an approximation to the cylindrical experiments that motivate the work, valid if the transport effects occur in a narrow region near the initial interface, which we have found to be true. We assume that all quantities only vary in the $\hat{\mathbf{x}}$ direction and that we have a magnetic field in the $\hat{\mathbf{z}}$ direction. This means that we are only interested in the transverse electric field and current components $E_y$ and $j_y$. This is shown in figure \ref{fig:geometry_diagram}. We also assume quasineutrality and solely consider hydrogen plasma with $Z = 1$ and $A = 1$, though this work can easily be generalised to any species. We ignore radiation losses and take a single-temperature approximation, so that we can replace the separate ion and electron energy equations with a single expression for plasma temperature. This assumption is further explored in section \ref{sec:temp_sep}. We use Braginskii fitting functions for the transport coefficients \cite{braginskii}, though changing to a different fitting function is a trivial change in the method. Recent work \cite{sadler_2021, epperlein_haines} has cast doubt on the accuracy of the fits to these transport coefficients, but the changes to the coefficients relevant in our geometry are minor and do not qualitatively change our conclusions. We use Gaussian units throughout, with temperature in energy units.

\subsection{Governing equations} 

The governing equations for our system are conservation equations for number density, momentum and energy. We solve these together with Faraday's and Amp\`{e}re's laws for the electromagnetic fields. We show the form of these equations in our geometry below.

The continuity equation is given by:
\begin{align} \label{eq:continuity}
    \frac{\partial N}{\partial t} + \frac{\partial}{\partial x}(N u_x) = 0,
\end{align}
where $N$ is the ion number density and $u_x$ the centre of mass velocity. We note that due to quasineutrality we have $N = N_i = N_e$ for $Z = 1$. 
As we are studying subsonic phenomena, pressure waves will equilibrate any imbalances in the system on shorter timescales than are of interest. Therefore, we can replace our momentum equation with an equation for pressure balance. We make no assumption about the ratio of the magnetic and thermal pressures and so retain the magnetic pressure contribution:
\begin{align} \label{eq:p_balance}
    2NT + \frac{B^2}{8\pi} = \mathrm{const.} 
\end{align}
It is worth noting that there is no magnetic tension in this planar geometry. We have a single equation for energy balance, which we get by summing the electron and ion equations from Braginskii \cite{braginskii}. In the 1D geometry that we are considering, all heat conduction is cross-field and we have no Righi-Leduc effect. Thus, the energy equation is given by: 
\begin{align} \label{eq:t_balance}
    \frac{\partial}{\partial t}(3NT) + \frac{\partial}{\partial x}(3NT u_x) + 2NT\frac{\partial u_x}{\partial x} = \frac{\partial Q}{\partial x} + \frac{1}{4\pi}\frac{\partial B_z}{\partial x}E_y,
\end{align}
where
\begin{align} \label{eq:q}
    Q = \frac{NT\tau_{e}}{m_e}\bigg(\gamma_\perp^e + \sqrt{\frac{2m_e}{m_i}}\gamma_\perp^i\bigg)\frac{\partial T}{\partial x} + \frac{c\beta_\wedge T}{4\pi e} \frac{\partial B_z}{\partial x}
\end{align}
is the magnetised heat flux. $Q$ is comprised of thermal conduction, where we include both the electron and ion conductivities, and the rarely included Ettingshausen effect. The Ettingshausen effect is heat flux driven by the current in the $\mathbf{j}\times\mathbf{B}$ direction and forms a large part of our work in this study. The second term on the right hand side of the energy equation is the Ohmic heating contribution, representing the conversion of magnetic energy into the thermal energy of the plasma. $\gamma_\perp^{i,e}$ and $\beta_\wedge$ are dimensionless Braginskii fitting functions \cite{braginskii} to the transport coefficients. Their form as a function of $\omega_e\tau_e$, the electron Hall parameter, can be found in appendix \ref{sec:transport_coeffs}. $\tau_e$ is the electron collision time, given by:
\begin{align}
    \tau_e = \frac{3\sqrt{m_e}T^{3/2}}{4\sqrt{2\pi}e^4 Z N\mathrm{ln}\Lambda},
\end{align}
where $\mathrm{ln}\Lambda$ is the Coulomb logarithm, which we take in this work to be:
\begin{align}
    \ln\Lambda = \ln\bigg(\frac{T}{e\hbar}\sqrt{\frac{3m_e}{\pi N}}\bigg).
\end{align}
$\omega_e$ is the electron cyclotron frequency, given by:
\begin{align}
    \omega_e = \frac{eB}{m_e c}.
\end{align}
We have noticed that results can be sensitive to the choice of function for the Coulomb logarithm; this should be borne in mind when using the self-similar solutions in this paper.
We use the generalised Ohm's law to obtain the electric field. Starting from the form given in Braginskii \cite{braginskii}:
\begin{align}
    \mathbf{E}^* + \mathbf{u}\times\mathbf{B} = \frac{m_e}{ne^2}\frac{\partial \mathbf{j}}{\partial t} - \frac{1}{ne}\nabla\cdot\underline{\underline{P_e}} + \frac{\mathbf{j}\times\mathbf{B}}{en_e} + \frac{1}{en_e}\mathbf{R_e} \\ \nonumber
    + \frac{m_e}{en_e}\nabla\cdot \bigg[n_e(\mathbf{u_i}\mathbf{u_i}-\mathbf{u_e}\mathbf{u_e})\bigg],
\end{align}
where $\mathbf{u_e}$ and $\mathbf{u_i}$ are the electron and ion velocities respectively, $\mathbf{j}$ is the current density, $\underline{\underline{P_e}}$ is the electron pressure tensor and $\mathbf{R_e}$ is the rate of change of electron momentum due to collisions with ions. An exact form for $\mathbf{R_e}$ can be found in \cite{braginskii}. Here we follow Braginskii's notation of writing the comoving electric field without asterisk as $\mathbf{E}$ and the electric field in the lab frame as $\mathbf{E}^*$. We neglect electron inertia and viscosity effects, replacing the pressure tensor with a scalar electron pressure $p_e$. In our geometry, $\mathbf{u}$ only has an $\hat{x}$ component and $\mathbf{B}$ only has a $\hat{z}$ component. The electric field is only used to evolve the magnetic field through Faraday's law $\partial_t \mathbf{B} = -c \nabla\times\mathbf{E^*}$. Therefore, only the transverse $\hat{y}$ components of the electric field are relevant for our case. This means that we have no Hall or Biermann effects. We can therefore write our generalised Ohm's law as:
\begin{align} \label{eq:E_full}
    E_y^* - \frac{1}{c}u_xB_z = E_y,
\end{align}
where $E_y$ is the frictional component of the electric field (the $\hat{y}$ component of $\mathbf{R_e}$), given by:
\begin{align} \label{eq:E}
E_y = -\frac{c m_e \alpha_\perp}{4\pi e^2 N \tau_{e}}\frac{\partial B_z}{\partial x}  -\frac{\beta_\wedge}{e}\frac{\partial T}{\partial x}.
\end{align}
The first term is the electric field set up due to diffusion of charge carriers down magnetic field gradients, whilst the second is the electrothermal Nernst effect. $\alpha_\perp$ is again a dimensionless fitting function given in appendix \ref{sec:transport_coeffs}. By taking the curl of the electric field and substituting into Faraday's law $\partial_t B_z = -\partial_x E_y^*$, we obtain the induction equation:
\begin{align} \label{eq:induction}
    \frac{\partial B_z}{\partial t} + \frac{\partial}{\partial x}(u_x B_z) = -c\frac{\partial E_y}{\partial x}.
\end{align}
This system of equations (\ref{eq:continuity}), (\ref{eq:p_balance}), (\ref{eq:t_balance}), (\ref{eq:E}) and (\ref{eq:induction}) is completed with boundary conditions on all the non-auxiliary variables $N, T, B$ and $u$. In this work, we consider a system divided into two uniform half-spaces for $t \leq 0$. We make no assumptions about the values of the variables in those half spaces, leaving them completely general. This leads to the initial conditions: 
\begin{align} \label{eq:BCs}
    T(x = -\infty, t = 0) &= T_0 & T(\infty, 0) &= T_1 \nonumber\\ 
    N(-\infty, 0) &= N_0 & N(\infty, 0) &= N_1 \nonumber \\ 
    B_z(-\infty, 0) &= B_0 & B_z(\infty, 0) &= B_1 \nonumber \\ 
    u_x(-\infty, 0) &= 0 & u_x(\infty, 0) &= 0,
\end{align}
At $t = 0$, the rigid thin interface at $x = 0$ is removed and the plasma in the half-spaces is allowed to evolve. The self-similar solutions, which this paper focuses on, represent the late-time asymptotic behavior of the system which should not depend on its exact initial state \cite{barenblatt}. However, the asymptotic behavior does depend crucially on the system states at plus and minus infinity. One should expect that far enough from the original interface, the plasma parameters are still given by eq. (\ref{eq:BCs}), so in our self-similar analysis this equation serves as boundary conditions. In our full hydrodynamic simulations, eq. (\ref{eq:BCs}) are taken as the actual initial conditions, but as such only affect the transition to the self-similar regime. 

It should be noted that in this section and the following ones, we use the Eulerian form of the governing equations. A comparison with the Lagrangian form, used in Garanin's previous work \cite{garanin}, can be found in the appendix. 
\subsection{Self-similar equations}
The 1D planar transport equations under the assumption of pressure balance are amenable to self-similar solutions. We define our self-similar variable:
\begin{align} \label{eq:ssv}
    \eta = \eta_0 \frac{x}{\sqrt{t}},
\end{align}
where $\eta_0$ is an arbitrary normalisation coefficient and introduce normalised self-similar variables:
\begin{align} \label{eq:normalisations}
    T(x, t) &= T_0 \theta(\eta) & N(x, t) &= N_0 n(\eta) \nonumber\\ 
    B_z(x, t) &= H_0 h(\eta) & Q(x, t) &= \frac{Q_0}{\sqrt{t}} q(\eta) \nonumber\\
    E_y(x, t) &= \frac{E_0}{\sqrt{t}} \epsilon(\eta) & u_x(x, t) &= \frac{u_0}{\sqrt{t}}v(\eta).
\end{align}
where $H_0 = \sqrt{16\pi N_0 T_0}$ is the normalisation using for the magnetic field, and $E_0$, $Q_0$ and $u_0$ are functions of the other normalisation variables with the correct physical dimension. These normalisation variables are all defined in appendix \ref{sec:normalisation_variables}.

As all of the spatial derivatives in equations (\ref{eq:continuity}) to (\ref{eq:induction}) are second order and the temporal derivatives are first order, this partial differential equation system scales in the same way as the diffusion equation. The ``diffusive scaling'' is what allows us to reduce the PDE system to ordinary differential equations in $\eta$ \cite{barenblatt}.

We rearrange the continuity equation (\ref{eq:continuity}) to obtain:
\begin{align} \label{eq:dvde}
    \frac{dv}{d\eta} = \frac{1}{n}\bigg(\frac{\eta}{2} - v\bigg)\frac{dn}{d\eta}.
\end{align}
Rearranging the equation for electric field, we obtain an expression for the gradient of the self-similar magnetic field $h$:
\begin{align} \label{eq:dhde}
    \frac{dh}{d\eta} &= \bigg[\frac{\mathcal{B}\alpha_\perp}{n\hat{\tau}}-\frac{\mathcal{A}\mathcal{C}\beta_\wedge^2}{n\hat{\tau}\gamma_\perp}\bigg]^{-1}\bigg(\epsilon - \frac{\mathcal{C}\beta_\wedge q}{n\theta\hat{\tau}\gamma_\perp}\bigg),
\end{align}
where $\mathcal{A} = cT_0H_o\eta_0/4\pi eQ_0, \mathcal{B}=cm_eH_0\eta_e/4\pi e^2 N_0 \tau_{e0} E_0$ and $\mathcal{C} = T_0 \eta_0 / e E_0$ are constant and dimensionless functions of the normalisation variables. $\gamma_\perp$ is the combined electron and ion conductivity: 
\begin{align}
    \gamma_\perp &= \gamma_\perp^e + \sqrt{\frac{2m_e}{m_i}}\gamma_\perp^i,
\end{align}
and $\hat{\tau}$ represents the variation of the electron collision time $\tau_e$ with the self-similar variable $\eta$:
\begin{align}
    \hat{\tau} &= \frac{\theta^{1.5}}{\tilde{\lambda}n},
\end{align}
where $\tilde{\lambda}$ is the function denoting the evolution of the Coulomb logarithm with $\eta$:
\begin{align}
    \tilde{\lambda} = 1 + \frac{1}{\ln\Lambda_0}\ln\bigg(\frac{\theta}{\sqrt{n}}\bigg).
\end{align} $\tau_{e0}$ and $\ln\Lambda_0$ are both given in appendix \ref{sec:normalisation_variables}.
The heat flux equation (\ref{eq:q}) can be rearranged for the self-similar temperature gradient:
\begin{align} \label{eq:dtde}
    \frac{d\theta}{d\eta} = \frac{1}{n\theta\hat{\tau}\gamma_\perp}\bigg(q - \mathcal{A}\beta_\wedge\theta\frac{dh}{d\eta}\bigg),
\end{align}
whilst the temperature balance equation is used to give an equation for the divergence of the heat flux:
\begin{align} \label{eq:dqde}
    \frac{dq}{d\eta} = 3n\bigg(v-\frac{\eta}{2}\bigg)\frac{d\theta}{d\eta} &+ 2n\theta \frac{dv}{d\eta} \nonumber \\ 
    -& \mathcal{A}\beta_\wedge\frac{dh}{d\eta}\frac{d\theta}{d\eta} - \frac{\mathcal{B}}{\sqrt{2\pi}}\frac{\alpha_\perp}{n\hat{\tau}}\bigg(\frac{dh}{d\eta}\bigg)^2
\end{align}
Finally, the induction equation is used to obtain an expression for the derivative of the electric field:
\begin{align} \label{eq:depsde}
    \frac{d\epsilon}{d\eta} = \mathcal{D}\bigg[-\frac{\eta}{2}\frac{dh}{d\eta} + v\frac{dh}{d\eta} + h\frac{dv}{d\eta}\bigg].
\end{align}
The density is not explicitly evolved, only being calculated from the normalised pressure balance:
\begin{align} \label{eq:ss_pbalance}
    n\theta + h^2 = 1 + \frac{1}{\beta_0},
\end{align}
where 
\begin{align} \label{eq:beta_0}
    \beta_0 = \frac{2n_0T_0}{B_0^2/8\pi} = \frac{H_0^2}{B_0^2}
\end{align}
is the ratio of the original thermal and magnetic pressures.
\subsection{Boundary conditions}
The equations are evolved as a 5D vector $\vec{y}(\eta) = (\theta, h, \epsilon, q, v)$, and the boundary conditions in equation (\ref{eq:BCs}) are instead converted to boundary conditions on the self-similar variables:
\begin{align} \label{eq:SS_BCs}
    \theta(-\infty) &= 1 & \theta(\infty) &= \frac{T_1}{T_0} \nonumber\\ 
    n(-\infty) &= 1 & n(\infty) &= \frac{N_1}{N_0} \nonumber \\ 
    h(-\infty) &= \frac{B_0}{H_0} = \sqrt{1/\beta_0} & h(\infty) &= \frac{B_1}{H_0} \nonumber \\ 
    v(-\infty) &= 0 & v(\infty) &= 0.
\end{align}
As we have 5 first order equations, we can only enforce 5 of these 8 boundary conditions. The density $n$, temperature $\theta$ and magnetic field $h$ are not independent variables due to the pressure balance, so it does not matter which one is chosen for the boundary conditions. In this work the density and magnetic field are both constrained at $\eta = -\infty$ and $\eta = \infty$ and the velocity at $\eta = -\infty$. This does mean that some of our solutions do not satisfy $v=0$ at $\eta = \infty$. Velikovich \textit{et al.} found that the velocity was equal to zero on both sides of the domain \cite{Velikovich_2019}. This is a consequence of the high $\beta$ assumption simplifying the energy equation considerably. In the low $\beta$ limit, the advection of magnetic pressure in the energy equation prevents the velocity being equal to zero on both sides in general \cite{Garcia-Rubio_20181}. In fact, in the zero $\beta$ limit \cite{garanin}, an infinite speed of energy supply from the source of magnetic flux is required to maintain pressure balance. We observe the same qualitative behaviour in our results, with an increasing $v(\eta=\infty)$ as the $\beta$ is decreased. 
\section{Method}
\label{sec:method}
Equations (\ref{eq:dvde}) to (\ref{eq:ss_pbalance}) constitute a system of nonlinear ordinary differential equations that need to be solved subject to the boundary conditions given by (\ref{eq:SS_BCs}). In this work, we solve these equations using a shooting method. We make a guess for the values of the variables $(\theta, h, \epsilon, q, v)$ at $\eta = 0$ and propagate the solution to $\eta = \pm \infty$ using an implicit ODE solver. We have experimented with several solvers, but have found that explicit ODE solvers are too costly for this stiff problem. The difference between the values of the variables at $\pm \infty$ and the desired boundary conditions is then calculated. Minimising this quantity to a specific tolerance completes the solution of this problem. Several approaches to this were tried, including Bayesian optimisation using the \texttt{GPyOpt} library and function minimisation using the \texttt{scipy.optimize} library, but have found root-finding using a simple Newton-Raphson method to be the most efficient and easiest to implement. 

The number of degrees of freedom and the nonlinearity of the equations mean that a shooting algorithm can be unstable. We have found that a poor guess for the value of the variables at the initial interface will generally lead to the final solution failing to converge. Unfortunately, this is a known problem of shooting methods and it is not easy to prevent. An effective mitigating strategy is to slowly move in parameter space allowing the method to converge at each step. For example, if changing the density boundary condition from $n(\infty) = N_1$ to $n(\infty) = N_2$, it should be changed from $N_1$ to $N_1+\alpha(N_2-N_1)$ where $\alpha$ is small. The size of this required step depends on the problem, but we have found that it needs to be smaller for highly magnetised problems. 

To demonstrate the utility of self-similar solutions for verification of MHD codes, we compare our results with a code that solves the full MHD equations without the assumption of pressure balance. For this, we use the Chimera radiation-MHD code, extensively used for laboratory astrophysics \cite{Chittenden_2004, ciardi_2007} and recently upgraded to include extended MHD effects \cite{walsh_2019, Walsh_2020_2}. We use the same initial and boundary conditions in Chimera as those used in the self-similar code. It is not possible to exactly satisfy pressure balance in the initial conditions in Chimera, due to the staggered grid used in implementation of the MHD algorithm. The initial discontinuity leads to compressive sound waves that propagate outwards from the initial interface. The region of the profile satisfying the self-similar solution is left in the wake of these pressure waves, so the comparison needs to be made after enough time has passed for the waves to move away from the region of interest. In testing, we have found that violations from pressure balance in the wake of the waves are less than 1\% and thus can be neglected.

\section{Results}
\label{sec:results}
\subsection{Low beta parameter scan} \label{sec:param_scan}
As discussed in the introduction, past work has focused on the high $\beta$ regime where the Ettingshausen and Ohmic heating terms can be dropped from the energy equation and the resistive diffusion term has a negligible impact in the induction equation. In this section, we perform a parameter scan over plasma $\beta$ to show how these neglected terms take effect and their qualitative roles as the $\beta$ is reduced. However, it is not possible to fully characterise the parameter space in terms of just the $\beta$ since as we will show the relative impact of different terms depends on the boundary conditions. Several sets of boundary conditions can have the same plasma $\beta$.

Instead, in this section, we refer to different sets of boundary conditions which we summarise in table \ref{table:BCs}. These conditions are intended as an illustrative set, which vary in plasma $\beta$ and plasma profile and indicate the expected behaviour in different situations. They are motivated by examples of MIF experiments. 
\begin{table*}
\centering
\begin{ruledtabular}
\begin{tabular}{@{}|l|ll|ll|ll|@{}}
 \textbf{Case} & N($x = \infty$)/cm\textsuperscript{-3} & N($x = -\infty$)& B($x = \infty$)/T & B($x = -\infty$) & T($x = \infty$)/eV & T($x = -\infty$) \\ 
 A & $7.5\times10^{23}$ & $1.5\times10^{23}$ & 1000 & 1000 & 5000 & 1000 \\
 B & $5\times10^{19}$ & $10^{20}$ & 150 & 0 & 250 & 100 \\
 C & $4.7\times10^{20}$ & $1.42\times 10^{22}$ & 2500 & 0 & 160 & 533 \\
 D & $10^{20}$ & $2\times10^{20}$ & 150 & 0 & 250 & 250 \\
 E & $4.7\times10^{20}$ & $4.78\times 10^{22}$ & 2500 & 0 & 160 & 160 \\
\end{tabular}
\end{ruledtabular}
\caption{Table of boundary conditions that will be discussed in the section.}
\label{table:BCs}
\end{table*}

\subsubsection{High beta plasma}

\begin{figure*}
    \centering
    \includegraphics[width = 1.0\textwidth]{"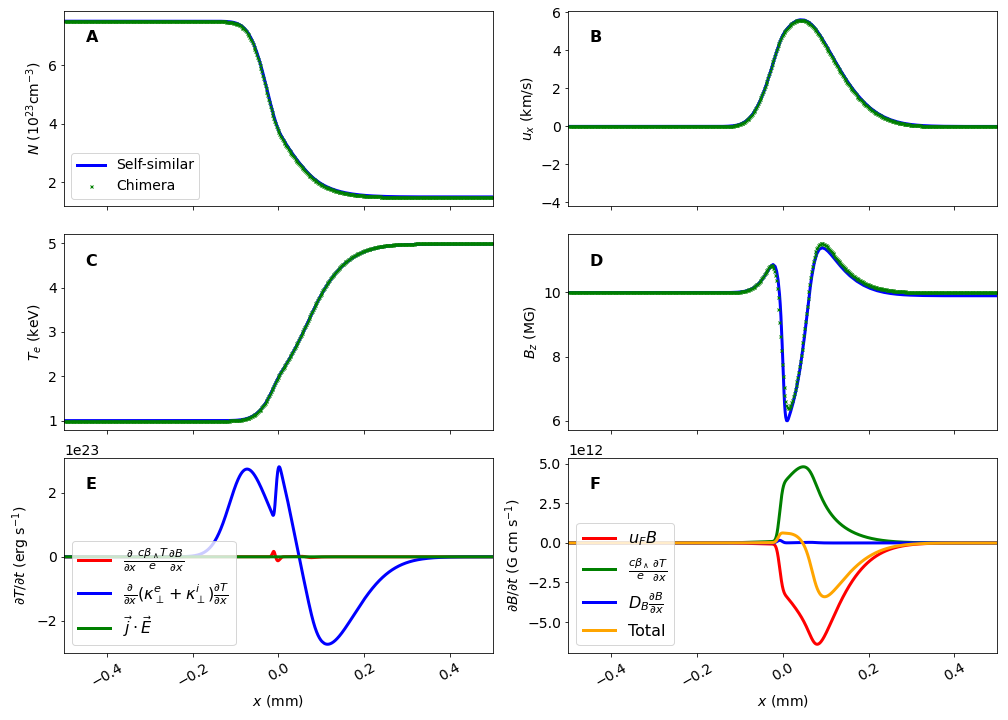"}
    \caption{\textbf{A-D}: Plots comparing the profiles calculated for the high $\beta$ boundary conditions (case A in table \ref{table:BCs}) for the self-similar code (in blue) to Chimera (in green). The profiles are compared after 4ns. \textbf{E}: The contribution to $\partial T/\partial t$ of the different terms in the energy equation (the Ettingshausen in red, the Ohmic heating in green, the Ohmic heating in blue) at the same time, calculated from the self-similar code. \textbf{F}: The contribution to $\partial B/\partial t$ of the different terms in the induction equation (the frozen-in-flow in red, the Nernst in green, the resistive diffusion in blue).}
    \label{fig:high_beta_ssvsChimera}
\end{figure*}

First, we consider the high $\beta$ case. We use the ``Nernst wave'' boundary conditions of Velikovich \textit{et al.} \cite{Velikovich_2019}, marked as case A in table \ref{table:BCs}. These are motivated by measured experimental conditions in MagLIF during stagnation. We calculate self-similar profiles using these boundary conditions and then compare to profiles calculated using the same boundary conditions with the Chimera code. We make the comparison after 4ns to allow the Chimera profiles to settle to pressure balance. This comparison is shown in panels A-D of figure \ref{fig:high_beta_ssvsChimera}. The initially discontinuous plasma is set into motion by thermal conduction; the hot plasma on the right hand side cools as energy is carried down the temperature gradient and thus contracts to increase density and maintain pressure balance. Similarly, the cold plasma on the left hand side heats and expands. This compression of the hot plasma can be seen by the peak of the velocity profile, which is to the right of the position of the initial discontinuity (at $x = 0$). The magnetic field does not significantly contribute to the pressure in this situation. Instead, its evolution is solely dictated by the balance between the Nernst, resistive diffusion and frozen-in-flow terms.  The left hand peak in the magnetic field is formed due to the Nernst effect advecting magnetic field down the temperature gradient - this is the Nernst wave described by Velikovich \cite{Velikovich_2019}. The right hand peak is due to the frozen-in-flow advection of the field by the plasma. In this situation, the resistive diffusion has a negligible effect. This can be seen from panels E and F of figure \ref{fig:high_beta_ssvsChimera}, which shows the contribution of the different transport terms to the energy and induction equations at 4ns (with the temperature and magnetic field profiles for comparison). Looking at the balance between the Ettingshausen, Ohmic heating and thermal conduction in the energy equation, it is clear that the heat conduction is hugely dominant over the other two terms and it is valid to neglect them in this case. In the induction equation balance, the Nernst and frozen-in-flow terms are significantly larger than the resistive diffusion. It should be noted that they carry magnetic flux in opposite directions in this situation. It can be seen that the global peaks in the Nernst and frozen-in-flow contributions to the induction equation are coincident with their corresponding peaks in the magnetic field profile. Whilst these terms are calculated at just a time snapshot, the self-similarity of our solution means that the plasma profiles broaden over time and thus the relative role of these gradient-driven transport terms remains qualitatively the same at later times (though the frozen-in-flow contribution decreases due to the velocity profile decreasing in magnitude at later times, see equation \ref{eq:normalisations}). These conclusions are consistent with previous theoretical work, but we have confirmed them self-consistently using our model \cite{Velikovich_2014, Velikovich_2019}. We have checked and solving these equations in an Eulerian (as here) or Lagrangian fashion (as in Garanin \cite{garanin}) yields identical results, as expected.

The agreement between the self-similar model and Chimera is good. There is some slight disagreement in the magnetic field profile. This is to be expected and is a consequence of the outward pressure waves as discussed in section \ref{sec:method}. This is also seen in other studies \cite{Velikovich_2019}. Though we use Chimera here as validation of our self-similar solution, this serves as a demonstration of how the self-similar code can be used as a test problem for MHD. Obtaining the same solution as the self-similar code requires correct implementations of frozen-in-flow, Nernst advection and thermal conduction - as well as normal hydrodynamic behaviour. Thus, obtaining agreement rigorously tests that the MHD code is solving the underlying equations correctly.

The initially uniform magnetic field in this case means that no steep magnetic field gradients are set-up and thus no large currents are induced. This reduces the impact of the Ettingshausen, Ohmic and resistive diffusion effects. However, we have found that even if there is a large gradient in the magnetic field (with 0T on the left and 1000T on the right), the Nernst, frozen-in-flow and heat conduction terms still dominate the transport in the high $\beta$ regime. The only way to increase the magnetic field gradient in our model is by increasing the field strength, we set the boundary conditions and the gradient is self-consistently calculated. Thus, increasing the field gradient is equivalent to decreasing the plasma $\beta$ on one side of the domain. We explore the effect of this in the next section.

\subsubsection{Moderate to low beta with a temperature gradient}
\begin{figure*}
    \centering
    \includegraphics[width = \textwidth]{"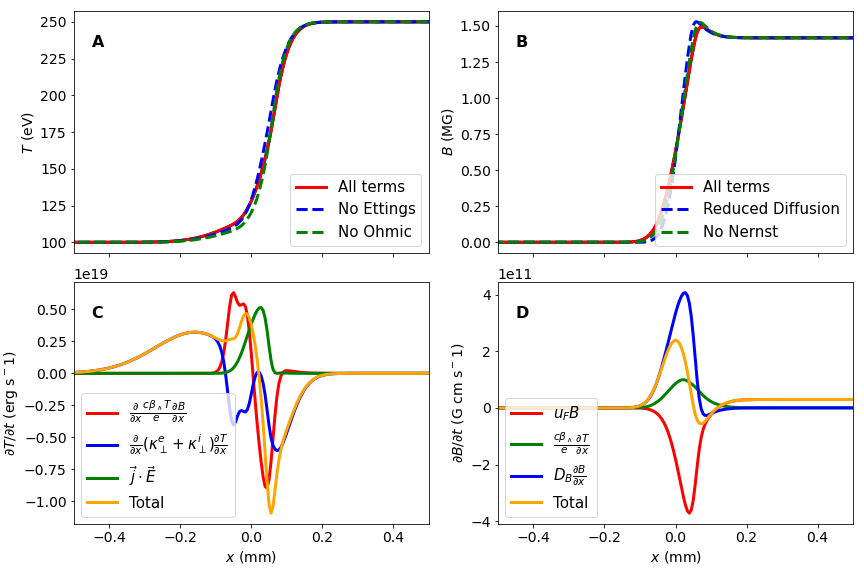"}
    \caption{\textbf{A}: Profile of the temperature profile calculated after a time of 4ns from the self-similar code for the initial conditions of case B in table \ref{table:BCs}. The profile is shown when all terms are included in the calculation (in red), when the Ettingshausen is excluded from the calculation (in blue) and when the Ohmic heating is not included (in green). \textbf{B}: The magnetic field profile with all terms included (in red), with the resistive diffusion coefficient reduced (in blue) and with the Nernst excluded (in green). \textbf{C}: The contribution of different terms to the energy equation, calculated from the self-similar code at 4ns. \textbf{D}: The contribution of different terms to the induction equation.}
    \label{fig:unity_beta_temp_gradient}
\end{figure*}

We now present results for a $\beta$ of order unity with a moderate temperature gradient, marked as case B in table \ref{table:BCs}. These parameters are indicative of what might be attained during the laser-preheat phase of MagLIF, although the field is significantly higher than in experiments. As before, we use our self-similar code to calculate profiles of density, temperature, magnetic field and velocity. The profiles of temperature and magnetic field are shown in panels A and B of \ref{fig:unity_beta_temp_gradient}. The competing magnetic transport processes are the Nernst effect and resistive diffusion carrying magnetic field down the temperature and magnetic field gradients respectively, from the right to the left. Both the Ettingshausen effect and the thermal conduction are carrying temperature in the same direction. The denser plasma on the left hand side is ablated as the magnetic field moves into it, expanding to reduce density and maintain pressure balance. This expansion of the plasma compresses the magnetic field by the frozen-in-flow effect, leading to the peak at about 0.08mm. At the same time, the damping of currents in the low density plasma leads to Ohmic heating, increasing the temperature. However, it is difficult to decouple which term is dominating here. The impact of ignoring certain terms in the calculation is shown in the temperature profile in figure \ref{fig:unity_beta_temp_gradient}. It can be seen that neglecting the Ettingshausen or the Ohmic heating changes the profile only slightly. In panel B, we show the effect on the magnetic field profile of neglecting particular terms. Our solution method does not allow setting the resistive diffusion to zero, so we show its qualitative effect by reducing the resistive diffusion coefficient by a factor of 2 and calculating the profiles. It is clear that both the Nernst and resistive diffusion effects have some impact on the structure of the field profile. It is certainly not the case that the Nernst effect is dominant, as seen in past work \cite{Giuliani_2018}. We also calculate the contributions to the induction and energy balance equations in panels C and D in figure \ref{fig:unity_beta_temp_gradient}. Looking at the balance of terms in the energy equation, it can be seen that the transport terms have a similar peak magnitude. However, the Ettingshausen and Ohmic heating terms act near the initial interface whilst the heat conduction acts over a larger region of space. This explains why the small impact of the Ettingshausen and Ohmic heating effects is confined to a small region in the temperature profile in figure \ref{fig:unity_beta_temp_gradient}. It also seems that in this case, the Ettingshausen and Ohmic effects have opposing effects on the temperature (see panel C at $x\approx0.7$mm). It should be noted that this is not generally the case and does depend on the magnetic field and temperature gradients, however it is discussed at length in section \ref{sec:temp_sep}. On the other hand, in the magnetic field balance the Nernst effect is significantly reduced compared to the resistive diffusion term, whilst the frozen-in-flow remains large. It can be seen from panel B that the Nernst effect essentially acts to increase the effective magnetic diffusivity of the plasma. This is observed in other work \cite{Velikovich_2014, Garcia-Rubio_20181}. 

We have tested the effect of having the temperature and magnetic field gradients in opposite directions, whilst maintaining the same unity plasma $\beta$ and have found the same qualitative behaviour of the transport terms. In this case, the Nernst opposes the resistive diffusion (effectively reducing the diffusivity) but it is still smaller in magnitude. The heat conduction acts over a longer scale and again the Ettingshausen and Ohmic heating somewhat counteract each other.

\begin{figure*}
    \centering
    \includegraphics[width = \textwidth]{"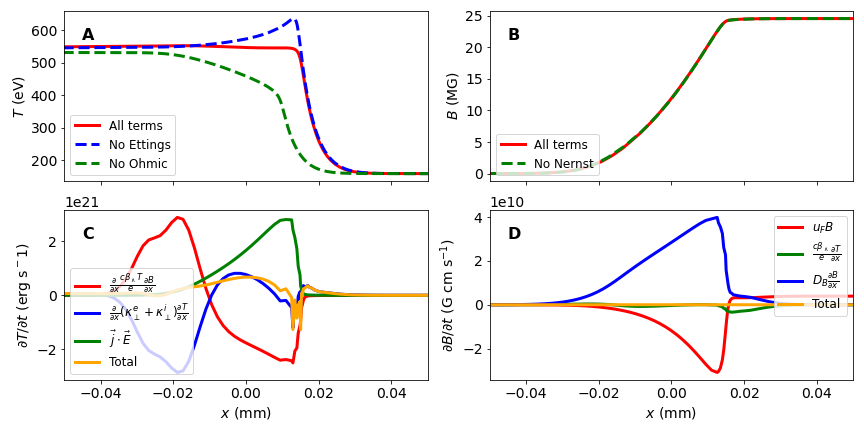"}
    \caption{\textbf{A}: Profile of the temperature profile calculated after a time of 4ns from the self-similar code for the initial conditions of case C in table \ref{table:BCs}. The profile is shown when all terms are included in the calculation (in red), when the Ettingshausen is excluded from the calculation (in blue) and when the Ohmic heating is not included (in green). \textbf{B}: The magnetic field profile with all terms included (in red) and with Nernst excluded (in green). \textbf{C}: The contribution of different terms to the energy equation, calculated from the self-similar code at 4ns. \textbf{D}: The contribution of different terms to the induction equation.}
    \label{fig:low_beta_temperature_gradient}
\end{figure*}

We then calculate the self-similar profiles for a low $\beta$ plasma, described by case C in table \ref{table:BCs}. In figure \ref{fig:low_beta_temperature_gradient}, we show the temperature and magnetic field profiles when particular terms are excluded from the calculation and the contribution of terms to the induction and energy equations. Looking at the terms in the induction equation, it is clear that the resistive diffusion is the dominant effect in magnetic field evolution across the entire domain and the Nernst effect has little impact. This is clear from the magnetic field profile that shows the Nernst having no effect. As the magnetic field diffuses into the dense plasma, magnetic energy is converted to thermal by Ohmic heating and the Ettingshausen effect advects energy down the magnetic field gradient. Neglecting the Ettingshausen effect in the calculation causes a noticeable peak in the temperature profile due to the absence of this additional advection. The increased temperature and magnetic pressure causes the dense plasma to expand to maintain pressure balance. Neglecting the Ohmic heating leads to less ablation of the dense plasma by the magnetic field, due to the reduced temperature and thus thermal energy. The leftward shift of the temperature profiles when Ohmic heating is excluded demonstrates this. Panel C shows the contribution of terms to the energy equation. It is clear that all terms have a comparable magnitude, with Ettingshausen and Ohmic heating again counteracting each other. The thermal conduction is balanced by the Ettingshausen on the left hand side of the interface. It is interesting that the Nernst effect cannot be observed whilst the Ettingshausen can, given that they both depend on the $\beta_\wedge$ transport coefficient. However, this is consistent with the theory that the ratio of the magnitudes of the Ettingshausen to the heat conduction approximately varies inversely with plasma $\beta$, where the ratio of the Nernst to resistive diffusion varies directly \cite{Velikovich_2014}.  Through comparison of figures \ref{fig:high_beta_ssvsChimera} through \ref{fig:low_beta_temperature_gradient}, it is clear that the length scales of temperature and magnetic field also reduce as the plasma $\beta$ is reduced. In particular, the temperature gradient scale length, $L_T$ becomes comparable to $L_B$. It is not possible to decouple whether the increased effect of e.g. the Ettingshausen term is due to this change to the length scales or directly due to the plasma $\beta$. This does however validate our earlier assumption that the transport processes occur in a narrow region near the interface. 

We have found that, despite the plasma $\beta$, if the magnetic field is initially uniform (as in case A), the currents induced by the frozen-in-flow and Nernst advection of magnetic field are not sufficient to cause significant Ohmic heating or Ettingshausen advection of the temperature. We believe this is the reason why past studies that did not impose a magnetic field gradient (e.g. \cite{Giuliani_2018}) are unable to observe a large effect due to these terms. This further emphasises our argument that it is not possible to classify the dominance of these terms in dimensionless parameters: it is profile dependent. This can be seen most clearly from the unity $\beta$ case shown in figure \ref{fig:unity_beta_temp_gradient}. Despite the Ettingshausen and heat conduction being of similar magnitudes, the Ettingshausen has almost no effect on the temperature profile as it acts over a much shorter length scale. We can show qualitatively that reducing the plasma $\beta$ reduces the temperature length scale to become comparable to that of magnetic field, but it clearly depends on the boundary conditions in a way that is difficult to determine a priori.

The benefit of our model is that it can recreate this low $\beta$ regime that is inaccessible to other models \cite{Velikovich_2014, Garcia-Rubio_20182, Velikovich_2019}, but also obtain consistent results in the high $\beta$ limit. We have not thoroughly explored even smaller plasma $\beta$ values, but we would expect the qualitative trend to continue - with the Ettingshausen and particularly the Ohmic heating terms increasing in importance. 

\subsubsection{A test problem for Ohmic heating and the Ettingshausen effects}
In the preceding sections, we showed how the balance of the transport terms is related to the plasma $\beta$. Ohmic heating is important in pulsed power experiments due to the lower $\beta$ in the surface plasma. Our analysis here implies that the Ettingshausen term may be important under these conditions as well. Given the wide variety of reduced MHD codes used to model these experiments \cite{Chittenden_2004, seyler_2011}, we argue that a test problem for these terms would be a useful tool. That is the purpose of this section.

\begin{figure*}
    \centering
    \includegraphics[width = 1.0\textwidth]{"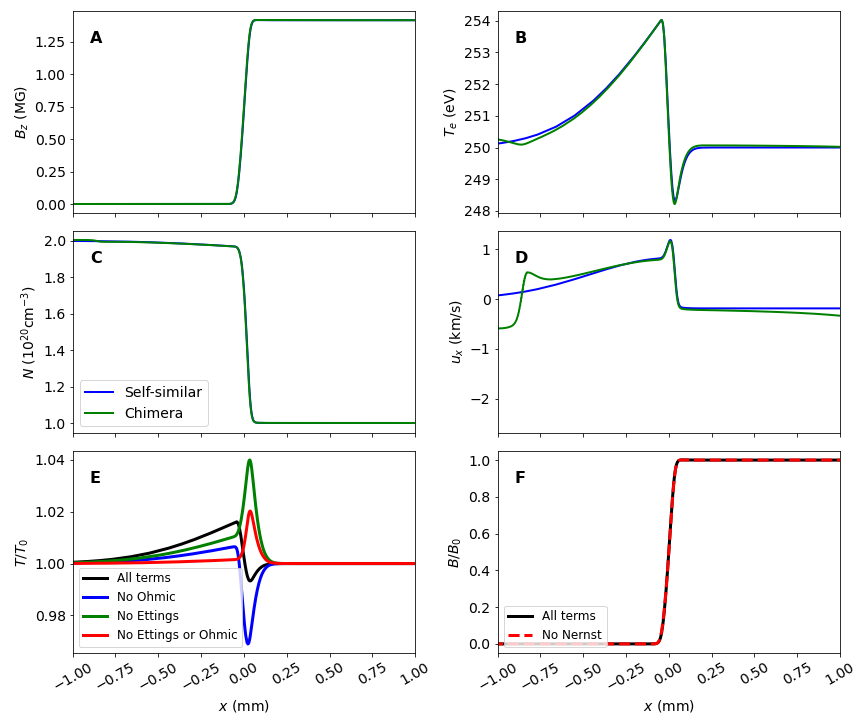"}
    \caption{\textbf{A-D}: The profiles of density, temperature, velocity and magnetic field for the order unity $\beta$ case, marked as case D in table \ref{table:BCs}. The results of the self-similar code are shown in blue, and the results using the MHD code Chimera are shown in green. The comparison is done at a time of 3ns. \textbf{E}: Temperature profiles when terms are excluded from the calculation. \textbf{F}: Magnetic field profile when the Nernst effect is excluded from the calculation.}
    \label{fig:unity_beta_comparison_with_chimera}
\end{figure*}

If we start with an initially constant temperature, this reduces the amount of thermal conduction and makes the role of the other transport terms more obvious. We use the boundary conditions of case D, given in table \ref{table:BCs}. These parameters are intended to be somewhat similar to the edge region of a z-pinch \cite{Chittenden_1993, Coppins_1992}. The comparison between our self-similar code and Chimera is shown in panels A-D in figure \ref{fig:unity_beta_comparison_with_chimera}. In panels E and F, we show the impact on the temperature and magnetic field profiles of ignoring particular terms in the calculation. The magnetic field profile is primarily determined by resistive diffusion into the dense plasma. This leads to Ohmic heating of the plasma due to the induced current and finite resistivity. As seen from panel E of figure \ref{fig:unity_beta_comparison_with_chimera}, Ohmic heating is responsible for the peak in the temperature profile. The same current drives advection of the temperature from right to left by the Ettingshausen effect. This leads to the formation of the trough in the temperature. Excluding both of these terms leads to a peak in the temperature due to pdV work, which can now dominate (see the red line in panel E of figure \ref{fig:unity_beta_comparison_with_chimera}). The agreement between the self-similar code and Chimera is good, with the exception of a pressure wave on the left hand side of the plot, as discussed in previous sections. This is therefore an effective test problem to test the implementation separately of the Ohmic heating and Ettingshausen terms. 

\subsection{Temperature separation} \label{sec:temp_sep}

\begin{figure*}
    \centering
    \includegraphics[width = 0.9\textwidth]{"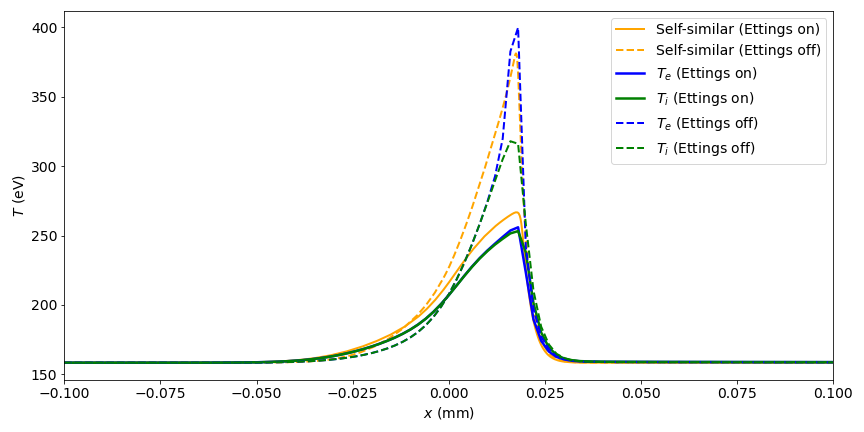"}
    \caption{Plot showing the temperature profiles produced by Chimera after 2.5ns for the magnetised interface problem, case E in table \ref{table:BCs}. The solid lines show the temperature calculated from Chimera with all terms included, the dashed lines show the profiles from Chimera with the Ettinghausen effect switched off. The orange lines show the solution calculated from our single-temperature self-similar code.}
    \label{fig:two_temp_simulation}
\end{figure*}

\begin{figure*}
    \centering
    \includegraphics[width = 0.9\textwidth]{"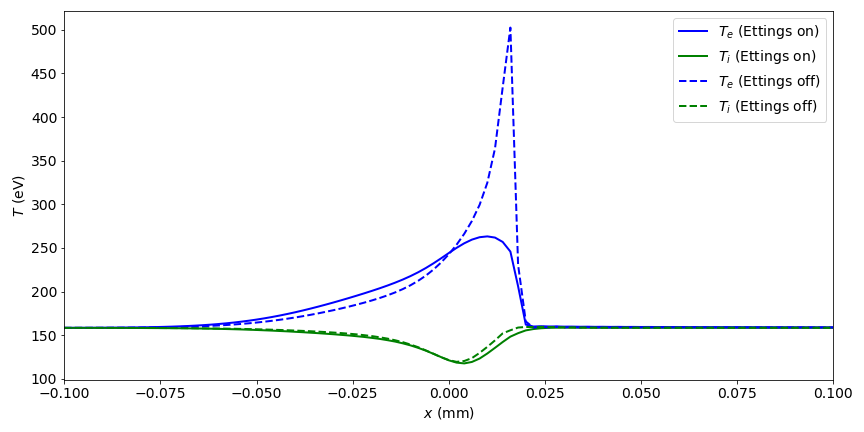"}
    \caption{Plots of the magnetised interface problem from Chimera after 2.5ns showing the electron temperature (in blue) and the ion temperature (in green) when the electron-ion equilibration term is set to zero. Solid lines include the Ettingshausen effect, dashed lines set it to zero.}
    \label{fig:no_equilibration}
\end{figure*}

We have assumed a single-temperature model in formulating our self-similar model. As the electron-ion energy equilibration term does not scale diffusively, it is not possible to develop a self-similar model that solves the two-temperature system of equations. Their lower mass means that the energy from Ohmic heating goes predominantly into the electrons, so where we have strong Ohmic heating and low density plasma, we may expect the electron and ion temperatures to separate. To model this effect, we solve the low $\beta$ boundary condition problem (defined by case E in table \ref{table:BCs}) using Chimera. This situation is chosen to be a proxy for a vacuum-plasma interface problem, and again is very similar to the edge region of a z pinch. The results of the Chimera simulation are shown in figure \ref{fig:two_temp_simulation}. We compare these results to the self-similar code. The assymmetry in the temperature profile is a consequence of the variation of the magnetisation. On the left, the plasma is unmagnetised, electron conductivity dominates and the temperature profile is broad. On the right, the plasma is highly magnetised and conductivity is reduced. While the agreement between Chimera and the self-similar solution is reasonable, it is challenging to get better agreement in this case as the small scale length of the temperature profile makes it difficult to get enough resolution in the region of interest whilst also leaving the simulation domain large enough for pressure waves to not affect the solution.
As predicted, the Ohmic heating causes the temperature to strongly peak. When the Ettingshausen effect is turned off (dashed lines), the electron and ion temperatures separate. The energy equilibration term is not sufficient to counteract the Ohmic heating. The magnitude of the Ohmic heating reduces later in time as the magnetic field gradient relaxes, but we have found that the temperature separation remains for 10s of nanoseconds. However, if the Ettingshausen effect is included in the calculation (solid lines), the temperatures do not significantly separate and both are substantially reduced. This is because the current responsible for Ohmic heating also drives strong advection of the temperature through the Ettingshausen effect. Figure \ref{fig:no_equilibration} shows the electron and ion temperatures when the equilibration term has been set to zero, i.e. with the temperatures decoupled from each other. It is clear that the ion profile is not directly affected by the Ettingshausen, but that the advection of the electron temperature by this effect is large. The Ettingshausen effect reduces the electron temperature sufficiently for the equilibration to bring it into balance with the ion temperature. It is likely that lower densities and higher magnetic fields would lead to stronger Ohmic heating and potentially temperature separation, but that even in this $\beta = 0.01$ situation, the Ettingshausen term is strongly mitigatory. This behaviour of the Ettingshausen effect has been observed before in z-pinch studies \cite{Chittenden_1993}. This suggests that in a situation where strong Ohmic heating is observed, neglecting the Ettingshausen effect could lead to overestimates of the electron and ion temperatures. In addition, this behaviour demonstrates that our self-similar solutions still produce quantitatively similar results to the two-temperature MHD code even despite the single-temperature assumption.

\section{Conclusions}
\label{sec:conc}
We have presented results from a new self-similar code that solves the 1D planar transport equations in a subsonic regime. This code is a generalisation of past work to arbitrary plasma $\beta$ and boundary conditions \cite{Velikovich_2014, Garcia-Rubio_20181, Garcia-Rubio_20182, Velikovich_2019}. The speed of this code makes it an ideal tool to assess the qualitative impact of various extended MHD terms, to decide whether they may be having an impact on a particular situation. In addition, as we have demonstrated in this work by comparison to the Chimera code, these self-similar solutions are effective test problems for MHD codes. Where Velikovich \textit{et al.} have developed the ``Nernst wave'' test problem for high-$\beta$ conditions, we believe that our test problem for the Ettingshausen and Ohmic heating terms could be useful to the wider MIF community, where further extended MHD terms may need to be included. Furthermore, these self-similar solutions are useful illustrations of extended MHD effects, aiding with qualitative understanding of these coupled and nonlinear phenomena.

We have shown that in high $\beta$ plasmas, using this isobaric model, the energy transport is dominated by thermal conduction and magnetic field transport by the Nernst term and frozen-in-flow. The picture is more complex in order unity $\beta$ plasmas. The relevant role of different MHD terms depends on the plasma profiles, certainly in the energy balance equation, and this cannot be parametised in terms of dimensionless numbers. We have found that for $\beta \approx \mathcal{O}(1)$ plasma with a temperature gradient, the heat conduction and Ettingshausen effects are of a similar magnitude, but the former acts over a larger spatial scale and thus the Ettingshausen effect has little impact on the temperature profile. In the low $\beta$ case, the Ohmic and Ettingshausen effects begin to dominate the temperature evolution but the thermal conduction cannot be neglected. For the magnetic field evolution, we have found that the resistive diffusion begins to dominate over the Nernst as the $\beta$ decreases to even order unity.

By suppressing thermal conduction by beginning with an initially uniform temperature, the qualitative impact of the Ettingshausen effect in advecting temperature down the magnetic field gradient becomes more apparent. This reduces the temperature peak and broadens the profile. This uniform temperature case can only be sensibly studied away from the high $\beta$ limit, where the magnetic field gradient can support the density gradient to maintain pressure balance.

We have also shown that the Ettingshausen effect inhibits temperature separation where there is strong Ohmic heating. This may mitigate high electron temperatures at vacuum-plasma interfaces (e.g. in \cite{Seyler_2018}) and suggests that the Ettingshausen should not be neglected if the Ohmic heating is found to be strong, such as in studies of z-pinches.

\begin{acknowledgments}
The authors would like to thank Dr Brian Appelbe for the use of his constant pressure code and Dr Alexander Velikovich for many helpful discussions.
\end{acknowledgments}

\section*{Data Availability Statement}
The data that support the findings of this study are available from the corresponding author upon reasonable request

\appendix

\section{Normalisation Variables}
\label{sec:normalisation_variables}
In section \ref{sec:theory}, we presented our ODE system of self-similar equations that we later solve. These self-similar equations are left in terms of dimensionless variables. This appendix defines the normalisation of these variables in terms of the physical parameters $N_0$ and $T_0$, the number density in $m^{-3}$ and temperature in energy units of our plasma. These variables are defined in equation (\ref{eq:normalisations}). \\ 
The self-similar variable is defined by $\eta = \eta_0\frac{x}{\sqrt{t}}$:
\begin{align}
    \eta_0 = \frac{N_0 T_0}{Q_0}.
\end{align}
$Q_0$ is the normalisation of the heat flux $Q(x, t)=Q_0 q(\eta)$, defined by:
\begin{align}
    Q_0 = \frac{N_0T_0^2\tau_{e0}(N_0, T_0)\eta_0}{m_e}.
\end{align}
The velocity is given by $u(x, t) = \frac{u_0}{\sqrt{t}}v(\eta)$, where
\begin{align}
    u_0 = \frac{1}{\eta_0}.
\end{align}
The electric field is $E(x, t) = \frac{E_0}{\sqrt{t}}\epsilon(\eta)$, where:
\begin{align}
    E_0 = \frac{\sqrt{8\pi}Q_0}{cH_0}
\end{align}
The collision time in terms of normalisation variables, given as $\tau_{e0}$ in section \ref{sec:theory} is given by:
\begin{align}
    \tau_{e0} = \frac{3\sqrt{m_e}T_0^{3/2}}{4\sqrt{2\pi}e^4 Z N_0\mathrm{ln}\Lambda_0},
\end{align}
and the Coulomb logarithm is given by 
\begin{align}
    \ln\Lambda_0 = \ln\bigg(\frac{T_0}{e\hbar}\sqrt{\frac{3m_e}{\pi N_0}}\bigg).
\end{align}
\section{Transport Coefficients}
\label{sec:transport_coeffs}
In section \ref{sec:theory}, we showed that transport coefficients are a key component of the equations that we solve in this work. Braginskii calculated dimensionless fitting functions to these transport coefficients \cite{braginskii} that we use in our calculations. These functions are functions of $\omega_e\tau_e$, the electron Hall parameter or magnetisation. They represent the anisotropic effect of magnetic fields on collisions. We repeat the functions here for completeness: 
\begin{align}
    \alpha_\perp = 1 - \frac{\alpha_1 \chi_e^2 + \alpha_0}{\chi_e^4 + \delta_1 \chi_e^2 + \delta_0},
\end{align}
where $\chi_e = \omega_e\tau_e$ is the electron Hall parameter, $\alpha_1 = 6.416$, $\alpha_0=1.837$, $\delta_1 = 14.79$ and $\delta_0 = 3.7703$. We also have
\begin{align}
    \beta_\wedge = \frac{\chi_e(\beta_1\chi_e^2 + \beta_0)}{\chi_e^4 + \delta_1\chi_e^2 + \delta_0},
\end{align}
where $\beta_1 = 1.5$ and $\beta_0 = 3.053$. $\gamma_\perp^e$ and $\gamma_\perp^i$ are defined by 
\begin{align}
    \gamma_\perp^e = \frac{\gamma_1 \chi_e^2 + \gamma_0}{\chi_e^4 + \delta_1 \chi_e^2 + \delta_0} \nonumber \\
    \gamma_\perp^i = \frac{2\chi_i^2 + 2.645}{\chi_i^4 + 2.6 \chi_i^2  + 0.677},
\end{align}
where $\chi_i = \omega_i \tau_i$ is the ion Hall parameter, $\gamma_1 = 4.664$ and $\gamma_0 = 11.92$.

\section{Comparison of Eulerian and Lagrangian equations}
\label{sec:appendix_lagrangian}
In section \ref{sec:intro}, we discuss the work of Garanin who solves the problem in the vacuum limit using a Lagrangian approach. Here we show how the approaches are related. The governing equations (\ref{eq:continuity}) to (\ref{eq:induction}) (see section \ref{sec:theory}) are recast into Lagrangian form, where 
\begin{align}
    \frac{d}{dt} = \frac{\partial}{\partial t} + u_x\frac{\partial}{\partial x}
\end{align}
in our geometry. Pressure balance remains the same, but the continuity equation becomes:
\begin{align}
\label{eq:continuity_lagrangian}
\frac{dN}{dt} + N\frac{\partial u_x}{\partial x} = 0.    
\end{align}
The energy equation is recast to:
\begin{align}
    \frac{d}{dt}(3NT) - 5T\frac{dN}{dt} = \frac{\partial Q}{\partial x} + \frac{1}{4\pi}\frac{\partial B_z}{\partial x}E_y,
\end{align}
where $Q$ and $E_y$ are defined in equations (\ref{eq:q}) and (\ref{eq:E}) respectively. Similarly, the induction equation becomes:
\begin{align}
\label{eq:induction_lagrangian}
    \frac{dB_z}{dt} - \frac{B_z}{N}\frac{dN}{dt} = -c\frac{\partial E_y}{\partial x}.
\end{align}
It should be noted that $dN/dt$ can be calculated from the pressure balance and so the continuity equation can be solved separately to the remainder of the equations. We introduce a new self-similar variable $\xi$, defined as:
\begin{align}
    \xi = \xi_0 \frac{\int_0^x N(X', t) dX'}{\sqrt{t}},
\end{align}
where $\xi_0$ is a normalisation constant. This obeys the transformations:
\begin{align}
    \frac{\partial}{\partial x} = \frac{\partial \xi}{\partial x}\frac{d}{d\xi} = \frac{\xi_0 N(x, t)}{\sqrt{t}}
\end{align}
and
\begin{align}
    \frac{\partial \xi}{\partial t} =& \xi_0 \frac{\int_0^x\frac{\partial N(X',t)}{\partial t} dX'}{\sqrt{t}} - \frac{\xi}{2t} \nonumber \\
    =& \xi_0\frac{\int_0^x\frac{\partial}{\partial X'}(NU)dX'}{\sqrt{t}} - \frac{\xi}{2t} \nonumber \\
    =& \xi_0\frac{N(x, t)U(x, t)}{\sqrt{t}} - \frac{\xi}{2t}.
\end{align}
Therefore:
\begin{align}
    \frac{d}{dt} = \bigg(\frac{\partial \xi}{\partial t} + u_x\frac{\partial \xi}{\partial x}\bigg)\frac{d}{d\xi} = -\frac{\xi}{2t}\frac{d}{d\xi},
\end{align}
which is obtained by using the continuity equation. We can apply these transformations to equations (\ref{eq:continuity_lagrangian}) to (\ref{eq:induction_lagrangian}) to obtain self-similar equations in terms of $\xi$. These are not shown here for brevity, but can be written as:
\begin{align}
    \frac{d\vec{y}}{d\xi} = \vec{F}(\vec{y}),
\end{align}
where $\vec{F}$ is a nonlinear operator and $\vec{y}(\xi) = (\theta, h, \epsilon, q)$. It is helpful to note that
\begin{align}
    \frac{\partial \xi}{\partial x} = \frac{\xi_0 N(x, t)}{\sqrt{t}} \nonumber \\
    \frac{\partial \xi}{\partial (\frac{x}{\sqrt{t}})} = \xi_0 N_0 n(\eta),
\end{align}
so
\begin{align}
\label{eq:dxide}
    \frac{\partial \xi}{\partial (\frac{1}{\xi_0 N_0}\frac{x}{\sqrt{t}})} &= n(\eta) \nonumber \\
    \frac{d\xi}{d\eta} &= n.
\end{align}
Therefore, instead of solving self-similar equations in $\xi$ and then having to find a non-trivial transformation back to $(x, t)$, we can transform our self-similar equations in $\xi$ to equations in $\eta$ and solve those. $v(\eta)$ can then be obtained by solving the continuity equation separately. We remark in section \ref{sec:results} that the same result is obtained either way, but that the Eulerian method offers improved numerical stability in the shooting method. \\
Now, we show that it is possible to transform between the self-similar variables $\eta$ and $\xi$ and thus that they are equivalent. If we assume
\begin{align}
    \xi = -n(v-\eta),
\end{align}
then
\begin{align}
    \frac{d\xi}{d\eta} =& -\frac{dn}{d\eta}(v-\eta) - n(\frac{dv}{d\eta} - 1) \nonumber \\
    =&-\frac{d}{d\eta}(nv) + \eta\frac{dn}{d\eta} + n \nonumber \\
    =& n,
\end{align}
as expected from equation (\ref{eq:dxide}). Therefore, we can easily convert between either set of self-similar equations. 
\nocite{*}
\bibliography{main}{}

\end{document}